\documentclass[showpacs,aps,floatfix,prb,reprint,superscriptaddress]{revtex4-1}
\usepackage{graphicx}
\usepackage{xfrac}
\usepackage{amssymb}
\usepackage{tikz}
\usepackage{url}
\usepackage{amsfonts}
\usepackage{amsmath}
\usepackage[utf8]{inputenc}
\usepackage{mathtools}
\usepackage{xcolor,colortbl}
\usepackage{bm}
\usepackage{array}
\usepackage{color}
\newcommand{\br}{\mathbf{r}}

\newcommand{\bk}{\mathbf{k}}
\newcommand{\bR}{\mathbf{R}}

\newcommand{\T}{\hat{\mathcal{T}}}

\newcommand{\bra}[1]{\langle #1 |}
\newcommand{\ket}[1]{| #1 \rangle}
\newcommand{\braket}[2]{\langle #1 | #2 \rangle}
\newcommand{\psit}{\tilde{\psi}}
\newcommand{\phit}{\tilde{\phi}}
\newcommand{\pt}{\tilde{p}}
\newcommand{\varphit}{\tilde{\varphi}}

\begin{document}

\title{Accurate Tight-Binding Hamiltonians for 2D and Layered Materials}
\author{Luis A. Agapito} 
\affiliation{Department of Mechanical Engineering and Materials Science, Duke University, Durham, NC 27708, USA}
\affiliation{Department of Physics, University of North Texas, Denton, TX 76203, USA}

\author{Marco Fornari}
\affiliation{Department of Physics, Central Michigan University, Mt. Pleasant, MI 48859}
\affiliation{Center for Materials Genomics, Duke University, Durham, NC 27708, USA}

\author{Davide Ceresoli}
\affiliation{CNR-ISTM, Istituto di Scienze e Tecnologie Molecolari. I-20133 Milano, Italy}

\author{Andrea Ferretti}
\affiliation{CNR-NANO S3 Center, Istituto Nanoscienze, I-41125 Modena, Italy}

\author{Stefano Curtarolo}
\affiliation{Center for Materials Genomics, Duke University, Durham, NC 27708, USA}
\affiliation{Materials Science, Electrical Engineering, Physics and Chemistry, Duke University, Durham, NC 27708, USA}

\author{Marco \surname{Buongiorno Nardelli}} 
\email{Email: mbn@unt.edu}
\affiliation{Department of Physics, University of North Texas, Denton, TX 76203, USA}
\affiliation{Center for Materials Genomics, Duke University, Durham, NC 27708, USA}
\date{\today}

\begin{abstract} 

We present a scheme to controllably improve the accuracy of tight-binding Hamiltonian matrices derived by projecting the solutions of plane-wave \textit{ab initio} calculations 
on atomic orbital basis sets. By systematically increasing the completeness of the basis set of atomic orbitals, we are able to optimize the quality of the band structure interpolation over wide energy ranges including unoccupied states. This methodology is applied to the case of interlayer and image states, which appear several eV above the Fermi level in materials with large interstitial regions or surfaces such as graphite and graphene. Due to their spatial localization in the empty regions inside or outside of the system, these states have been inaccessible to traditional tight-binding models and even to \textit{ab initio} calculations with atom-centered basis functions.

\end{abstract}
\maketitle 
\section{Introduction} 

The generation of highly accurate  tight-binding models for arbitrary systems is a long-lasting problem  that has enormous implications in the development of efficient tools for the study of the electronic structure of molecules and solids \cite{Huckel1931Model,Jones1934Tight_binding}, and for applications in accelerated materials development. \cite{nmatHT}
 With the introduction of \textit{ab initio} tight-binding Hamiltonians the accuracy of these methods has seen a substantial improvement. 
However, the best representations still rely on \textit{ad hoc} basis sets that need to be iteratively optimized \cite{Andersen_muffintin_PRB2000, Marzari2012MLWF, Lu2004QUAMBO} and are computationally expensive. 
In recent papers, we have introduced an efficient scheme to construct optimal tight-binding Hamiltonians projecting the Bloch states obtained from plane-wave (PW) Density Functional Theory calculations onto atomic orbitals derived directly from the generation of the atomic pseudopotentials.\cite{Agapito_2013_projectionsPRB,Agapito2015TB_minimalPAO,curtarolo:art93} In this scheme, the energy range in which the TB Hamiltonian reproduces the original states is limited by the finite number of pseudo atomic orbitals (PAO) that comprise the minimal basis set. As such, only a few unoccupied bands are typically well represented and an accurate description of the conduction states is impossible beyond a few eV. If more conduction states are needed, the basis set needs to be systematically extended.

In this work, we propose a procedure
that extends the validity of the TB representation of the band structure to electronic states far above the Fermi level.
This novel approach is  based on the Projector Augmented Wave (PAW) formalism and involves several atomic orbital (AO) for each angular momentum that are directly computed from the all electron atomic potential. 

The paper is organized as follow: in section II.A we introduce the PAW formalism; since the PAW method requires smoother AO functions, i.e. the PAOs, the pseudization process is presented in Section II.B; in Sec. II.C we briefly summarize the projection, filtering and shifting procedure to generate accurate Hamiltonian matrices as originally discussed in Ref. \onlinecite{Agapito_2013_projectionsPRB,Agapito2015TB_minimalPAO}; in Sec. II.D we discuss the convergence properties of the unoccupied  Kohn-Sham states and their dependence on the basis set representation; finally, in Section III we present two examples where we demonstrate the effectiveness of our enhanced scheme by reproducing the interlayer states of graphite and the image states of graphene in a wide energy range. 

\section{Methodology} 
\subsection{The PAW method}\label{sec:paw_method} 
Density-functional theory (DFT) in combination with the plane-wave
pseudopotential (PP) formalism is one of the most common method to derive
the electronic structure of molecules and solids. Pseudopotentials are
constructed to remove core electrons from the Hamiltonian and to reduce
drastically the number of plane waves that would be otherwise needed to
represent the divergent Coulomb potential close to the nucleus. In addition, 
pseudo-wavefunctions are smooth in the region close to the
nuclei because they don't need to be orthogonalized to the core orbitals.
The drawback of the PP method is that all the details on the 
wavefunctions within a specific distance from the nucleus (the atomic sphere)  are lost. This is important 
when it comes to calculate, for instance, magnetic resonance parameters or core emission
spectroscopies.\cite{Pavan201259,PhysRevB.66.195107,PhysRevB.53.10942}
In that regard, the PAW method\cite{Bloechl1990GeneralizedPotentials} allows
to reconstruct the full nodal structure of the wavefunctions near the ions.
The basic idea is to ``augment'' the pseudo wavefunctions with
a set of pseudo partial waves $\ket{\varphit_\alpha^n}$, which are localized functions centered at position $\bR_\alpha$
for each atom $\alpha$
($n$ is a composite index for the quantum
numbers $n,l,m$). This augmentation procedure is achieved through the
application of the $\T$ operator to the pseudo wavefunctions, $\ket{\psit_i}$,
in order to obtain the all-electron wavefunctions,  $\ket{\psi_i} = \T \ket{\psit_i}$.
Such an operator is defined as 
\begin{eqnarray}
  \T &=& \hat{1} + \sum_\alpha \T_\alpha,      \nonumber\\
  \T_\alpha &=& \sum_n \left( \ket{\varphi_\alpha^n} - \ket{\varphit_\alpha^n} \right)
         \bra{\tilde{p}_\alpha^n}
  \label{eq:PAW}
\end{eqnarray}
where 
$\ket{\varphi_\alpha^n}$ are the all-electron partial waves and $\ket{\varphit_\alpha^n}$ are
the corresponding pseudo partial waves. The PAW method is grounded
on the assumption of completeness of the basis of partial waves in Eq.~\ref{eq:PAW}: the
wavefunction can be expanded in terms of partial waves inside the
augmentation sphere. In practice, in order to enforce a high degree
of completeness one has to include more than one partial wave per angular
momentum (typically two or three), where the first corresponds to the
bound energy state and the others to unbound states of positive energy. 
The projectors $\ket{\tilde{p}_\alpha^n}$ are local functions centered at $\bR_\alpha$
 and vanishing beyond a certain cutoff radius $r_\alpha^c$; they are 
 determined such that:
\begin{equation}
\forall \alpha\  \braket{\tilde{p}_\alpha^n}{\varphit_\alpha^{n'}} = \delta_{n,n'} \quad
  \text{for} \quad |\br - \bR_\alpha| < r_\alpha^c
\end{equation}
The inclusion of multiple partial waves enables the high transferability
and accuracy of the PAW potentials, which are defined by a given set of partial waves, projectors and cutoff radii. Libraries of
PAW datasets for almost all elements in the periodic table are available:
PSlibrary\cite{[{}][{. \url{http://www.qe-forge.org/gf/project/pslibrary/}}]DalCorso2014Pseudos},
GBRV\cite{[{}][{. \url{http://www.physics.rutgers.edu/~dhv/gbrv/}}]gbrv}, 
JTH\cite{[{}][{. \url{http://www.abinit.org/downloads/PAW2}}]Jollet2014PAWlib}, 
GPAW\footnote{\texttt{https://wiki.fysik.dtu.dk/gpaw/setups/setups.html}}, 
VASP\cite{kresse_vasp_paw}, 
ATOMPAW\footnote{\texttt{http://users.wfu.edu/natalie/papers/pwpaw/newperiodictable/}},
, and GIPAW, used in the calculation of NMR shifts\cite{[{}][{. \url{https://sites.google.com/site/dceresoli/pseudopotentials}}]Ceresoli2010GIPAW}.

\subsection{Generation of the PAO basis sets}\label{sec:generation_pao}

For given energies $\epsilon_n$, the all-electron atomic-orbital functions $\ket{\phi_\alpha^n}$ are obtained  by solving the  Schr\"odinger equation:
\begin{equation}
  \left[ -\frac{1}{2}\nabla^2 + V_\alpha^{\textrm{AE}} \right] \ket{\phi_\alpha^n} =   
{\epsilon_n}\ket{\phi_\alpha^n}\label{eq:radial_eq}
\end{equation}
where $V_\alpha^{\textrm{AE}}$ is the screened all-electron potential found
by self-consistently solving the isolated atom at a reference electronic 
configuration. 
%
%
The corresponding smooth atomic-orbitals $\ket{\phit^n}$ (dropping the atom
index $\alpha$ for simplicity) are obtained by solving the implicit
equation: $\ket{\phi^n} = \T \ket{\phit^n}$, that is:
\begin{equation}
  \phit^n(r) = \phi^n(r) - \sum_m {\left[
     \varphi^m(r)-\varphit^m(r) \right] C_{m,n}} \label{eq:inverseT}
\end{equation}

where $C_{m,n}=  \braket{\tilde{p}^m}{\phit^n} =  \int_0^{r_c} \pt^m(r)^*\phit^n(r) r^2 dr$. Notice 
that we have include only the radial component
of the functions, \emph{i.e.} without the the angular
part, given by the spherical harmonics $Y_{lm}$.

Eq.~(\ref{eq:inverseT}) has the form of a Fredholm equation of the second
kind\cite{Kanwal2013linear} which can be reduced to a matrix equation
 defining $C = (I+A)^{-1} B$ where $I$, $A$ and $B$ are identity, square and rectangular 
matrices, respectively, with elements $A_{k,h} =  \int_0^{r_c} \pt^k(r)^* (\varphi^h(r) - \varphit^h(r)) r^2 dr$, $B_{k,m} = \int_0^{r_c} \pt^k(r)^* \phi^m(r) r^2 dr$. The indexes $k$ and $h$ runs
over the set of projectors and partial waves that define the PAW dataset, respectively, while
the indexes $n,m$ run over the full set of radial wavefunctions from
Eq.~(\ref{eq:radial_eq}).

\begin{figure}[htb]
\begin{center}
\includegraphics[width=0.99\columnwidth]{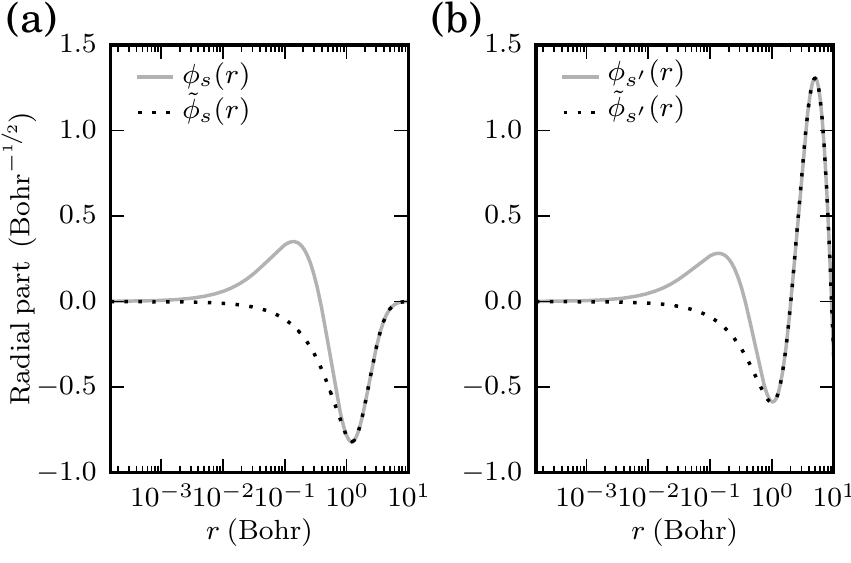}
\vspace{-5mm}
\caption{
All-electron and pseudo-atomic functions for the $\{s,s'\}$ components of the PAO$_3$ set. The parameters used to compute the pseudo-atomic orbitals are defined by the same PAW data set used in the DFT calculation.
}
\label{fig:AE_and_PS} 
\end{center}
\end{figure}

We have applied the procedure discussed above to derive 
sets of smooth AOs, \textit{i.e} the PAOs $\phit_i$. In a strict sense, while the PAOs are the auxiliary functions involved in the PAW method, the AOs $\phi_i$ are the functions that constitute the basis set for the tight binding models.
Fig.~\ref{fig:AE_and_PS} shows a comparison of the radial parts of the all-electron vs. the pseudo atomic orbitals of carbon (the $s$ and $s'$ functions of PAO$_3$ given in Table~\ref{tab:table1}) using the PAW potential \texttt{C.pbe-n-kjpaw\_psl.1.0.0.UPF} from Ref.~\onlinecite{[{}][{. \url{http://www.qe-forge.org/gf/project/pslibrary/}}]DalCorso2014Pseudos}. The PAO functions are smoother that the all-electron functions inside the cutoff radius of 1.4 $a_B$, while, by construction, both are identical outside the cutoff radius. 

One point of strength of this approach is that we
can construct PAO sets of increasing size and completeness. The sets can include multiple functions for each $\{lm\}$ channel which correspond to different energy parameters $\epsilon_n$ (see Table~\ref{tab:table1}). 
The first choice for the values of $\epsilon_n$ are the eigenenergies of the bound states ($\epsilon_n < 0$) of the isolated atom. The orbitals form the minimal set PAO$_1$ which is equivalent to the single-zeta basis used in quantum chemistry, \textit{e.g.} $\{s,p\}$ functions for carbon, composed of $M$=4 functions. 

\begin{table}[h]
\caption{\label{tab:table1}
Energy parameters (in Ry) that define the functions used in the construction of the PAO sets $\{\phit_n(\br)\}$ for carbon via Eq.~\ref{eq:radial_eq}.
} 
\begin{ruledtabular}
\begin{tabular}{c|ccc|cc|cc}
$\textrm{set}$ & $\epsilon_s$ & $\epsilon_p$ & $\epsilon_d$ & $\epsilon_{s'}$ & $\epsilon_{p'}$ & $\epsilon_{s''}$ & $\epsilon_{p''}$  \\ \hline
PAO$_1$ &-1.01 &-0.39 & --   &--  & --  & -- & -- \\ 
PAO$_2$ &-1.01 &-0.39 & 0.05 &--  & --  & -- & -- \\
PAO$_3$ &-1.01 &-0.39 & 0.05 &0.05& 0.05& -- & -- \\
PAO$_4$ &-1.01 &-0.39 & 0.20 &0.20& 0.20&0.4 &0.4 \\
\end{tabular}
\end{ruledtabular}
\end{table}

Hamann\cite{Hamann1989Generalized_NCPP} showed that positive-energy (scattering) states can be employed to improve the accuracy of norm conserving pseudopotentials. Within our approach, we augment the minimal PAO$_1$ set with scattering states of energies $\epsilon_n > 0 $ (these energy parameters can be chosen arbitrarly). 
The PAO$_2$ set is of single-zeta-polarized quality that includes polarization functions with higher angular momentum, $l_{\textrm{max}}+1$, \textit{i.e.} $\{s,p,d\}$, $M$=9. PAO$_3$ is of double-zeta-polarized quality, \textit{i.e.} $\{s,s',p,p',d\}$, $M$=13, and PAO$_4$ triples the number of minimal basis functions and includes additional polarization functions, \textit{i.e.} $\{s,s',s'',p,p',p'',d\}$, $M$=17, thus, it is of triple-zeta-polarized quality.

\subsection{Building the TB Hamiltonian matrices}\label{sec:building_matrxi} 

Accurate TB Hamiltonian matrices can be built from the direct projection of the Kohn-Sham (KS) Bloch states $\ket{\psi_{n\bk}}$ onto a chosen PAO set as discussed extensively in Ref.~\onlinecite{Agapito_2013_projectionsPRB,Agapito2015TB_minimalPAO}.
This procedure is satisfactory when Bloch states, that project well  on the selected AO basis set, are kept and states that do not project well are eliminated, 
\textit{i.e.}, filtering. In this process the crucial quantities that quantify the accuracy of the basis are the projectabilities $p_{n\bk}=\bra{\psi_{n\bk}} \hat{P} \ket{\psi_{n\bk}} \ge 0$ ($\hat{P}$ is the operator that projects onto the space of the PAO basis set, as defined in 
Ref.~\onlinecite{Agapito2015TB_minimalPAO}
) which indicate  the representability of a Bloch state $\ket{\psi_{n\bk}}$ on the chosen PAO set.
Maximum projectability, $p_{n\bk}= 1$, indicates that the particular Bloch state can be perfectly represented in the chosen PAO set; contrarily, $p_{n\bk} \approx 0$ indicates that the PAO set is insufficient and should be augmented. 
Once the Bloch states with good projectabilities have been identified, the TB Hamiltonian is constructed as: 
\begin{equation}
H(\bk) = AEA^\dagger + \kappa \left( I-A \left( A^{\dagger}A \right)^{-1}A^\dagger \right)\, . 
\label{eq:Hk}
\end{equation}
where $E$ is the diagonal matrix of KS eigenenergies and $A$ is the matrix of coefficients obtained from projecting the Bloch wavefunctions onto the PAO set (See Ref.~\onlinecite{Agapito2015TB_minimalPAO}.) Since the filtering procedure introduces a null space, the parameter $\kappa$ is used to shift all the unphysical solutions outside a given energy range of interest.

The real-space TB matrix, $H(\bR)$, between the central and the neighboring unit cell at lattice vector $\bR$ is obtained via Fourier transform: 
\begin{equation} 
H(\bR)=\frac{1}{N_{\mathcal{V}}} \sum\limits_{\mathbf{k}} e^{-i\mathbf{k}\cdot\mathbf{R}}H(\bk)\, ,
\label{eq:HR}
\end{equation}
where $N_\mathcal{V}$ is the number of $\bk$ points in the reciprocal unit cell. Using these matrices, one can calculate the interpolated TB band structure, for any $\mathbf{k}$, using the inverse Fourier transform.

\subsection{TB representation of the unoccupied bands}

When a linear combination of AOs (LCAO) are used as basis sets in DFT based methods, the unoccupied bands tend to substantially depend upon basis set size.
As an illustration of the above argument, we have computed the band structure of graphite and graphene using both approaches.
All PW- and LCAO-DFT calculations presented in this work were performed using the software packages \textsc{quantum espresso}\cite{quantum_espresso_2009} or \textsc{openmx}\cite{[{}][{. \url{http://www.openmx-square.org/}}]Ozaki2003BasisFunctions} using the Perdew-Burke-Ernzerhof (PBE)\cite{PBE} exchange and correlation functional.
The PW calculations use the PAW pseudopotential \texttt{C.pbe-n-kjpaw\_psl.1.0.0.UPF}  from the PSlibrary\cite{[{}][{. \url{http://www.qe-forge.org/gf/project/pslibrary/}}]DalCorso2014Pseudos} and a kinetic energy cutoff energy of 40 Ry. The LCAO-DFT calculations use norm conserving pseudopotentials and the optimized AO basis sets from the  \textsc{openmx} pseudopotential database\cite{[{}][{. \url{http://www.jaist.ac.jp/~t-ozaki/vps_pao2013/}}]Ozaki2004Adpack}, with a cutoff radius for all carbon basis functions of 7 $a_B$. The basis set used for the ``empty atoms'' contains two $s$, two $p$, two $d$, and one $f$ function, all with a cutoff of 13 $a_B$.

\begin{figure}[htb]
\begin{center}
\includegraphics[width=0.99\columnwidth]{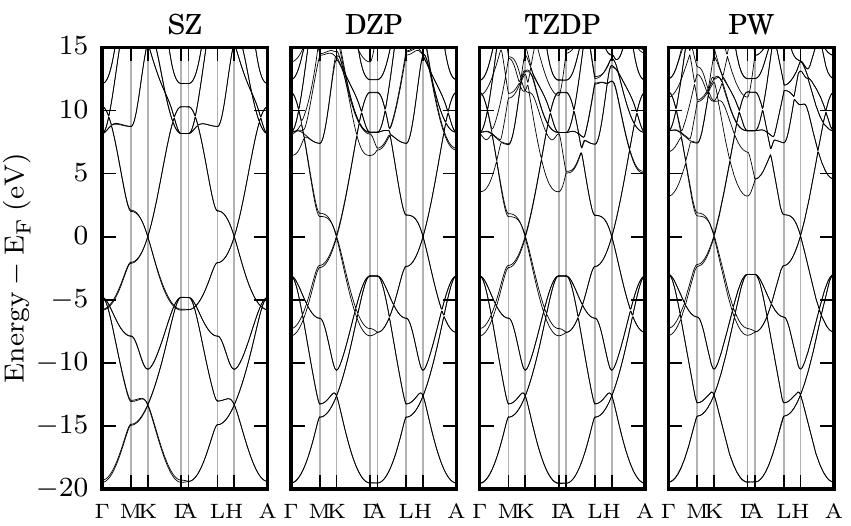}
\vspace{-5mm}
\caption{\small
Band structure of graphite using atomic-orbital (at the single-zeta, double-zeta polarized, and triple-zeta doubly-polarized level) and well converged plane-wave basis sets. 
}
\label{fig:convergence} 
\end{center}
\end{figure}

\begin{figure}[htb]
\begin{center}
\includegraphics[width=0.99\columnwidth]{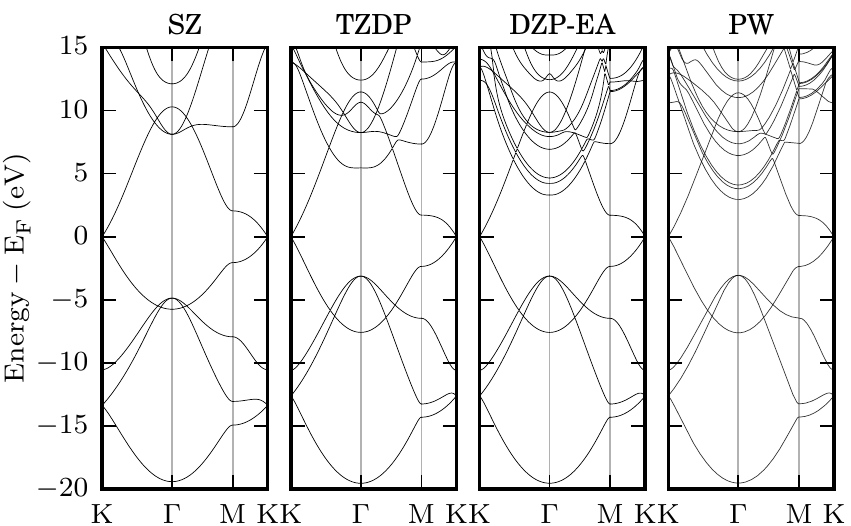}
\vspace{-5mm}
\caption{
Band structure of graphene using atomic-orbital (at the single-zeta, double-zeta polarized, and triple-zeta doubly-polarized level) and well converged plane-wave basis sets. The first three panels use atomic-orbital-like sets: SZ, TZDP, and DZP with empty atoms from Ref.~\onlinecite{Ozaki2004Adpack}.}
\label{fig:fig_graphene1} 
\end{center}
\end{figure}

The first three panels in Figs.~\ref{fig:convergence} and \ref{fig:fig_graphene1} show the LCAO-DFT band structures for graphite and graphene, respectively, calculated using AO basis sets of increasing quality, taken from a public database\cite{[{}][{. \url{http://www.jaist.ac.jp/~t-ozaki/vps_pao2013/}}]Ozaki2004Adpack}: single zeta (SZ), double zeta with polarization (DZP), and triple zeta with double polarization (TZDP). Under the same approximation to the exchange-correlation functional, calculations using a well converged plane wave (PW) basis set (fourth panel) reproduce the unoccupied states systematically better than the AO basis sets.

Not surprisingly, a minimal basis set such as SZ fails to reproduce not only the unoccupied but also the occupied bands of graphite and graphene. The SZ calculation completely misses the lowest conduction band at $\Gamma$, as seen in the first panel in Figs.~\ref{fig:convergence} and \ref{fig:fig_graphene1}.
DZP basis sets are generally considered satisfactory to reproduce ground-state properties, reaching close to chemical accuracy.\cite{Gusso2008LCAOvsPW} Indeed, we find that all occupied states are well converged at the DZP level; however, it offers little improvement to the unoccupied bands. Only when the much larger TZDP set is used, the unoccupied bands start to qualitatively match the fully converged PW results.

Despite the deficiencies of standard implementation of DFT, single particle KS eigenstates (occupied and unoccupied) are often needed, for instance, as the starting point for more refined calculation of the excited states (time-dependent DFT,\cite{Gross1996TDDFT} density-functional perturbation theory,\cite{Gorling1995DFPT} many-body perturbation theory GW,\cite{Hybertsen1986GW} coupled-cluster theory,\cite{Taube2008CCbasis} {\it etc.}), thus TB Hamiltonians that are expressed in small AO basis sets but that can still deliver the accuracy of the converged PW DFT results, especially for the unoccupied states, are a valuable tool for the study of novel materials and further development of theoretical methods.

\section{Applications}
For most crystalline materials, minimal PAO sets (of SZ or SZP quality) are sufficient for constructing TB Hamiltonians that are describing accurately the band structre up to $\sim 2$ eV above the Fermi energy.\cite{Agapito_2013_projectionsPRB,Agapito2015TB_minimalPAO} However, materials containing extended interstitial regions tend to exhibit ``interlayer states'' usually located several electron-volts above the Fermi energy.\cite{} Similarly, the description of ``image states'' in metallic surfaces\cite{} also requires a larger energy range in the unoccupied bands. Here, we discuss two prototypical cases, graphite and graphene, to demonstrate the effectiveness of our augmented PAO basis sets and TB scheme for the treatment of interlayer and image states, which have been out of reach of traditional parameterized TB models so far.

\subsection{Graphite}
\begin{figure}[htb]
\begin{center}
\includegraphics[width=0.99\columnwidth]{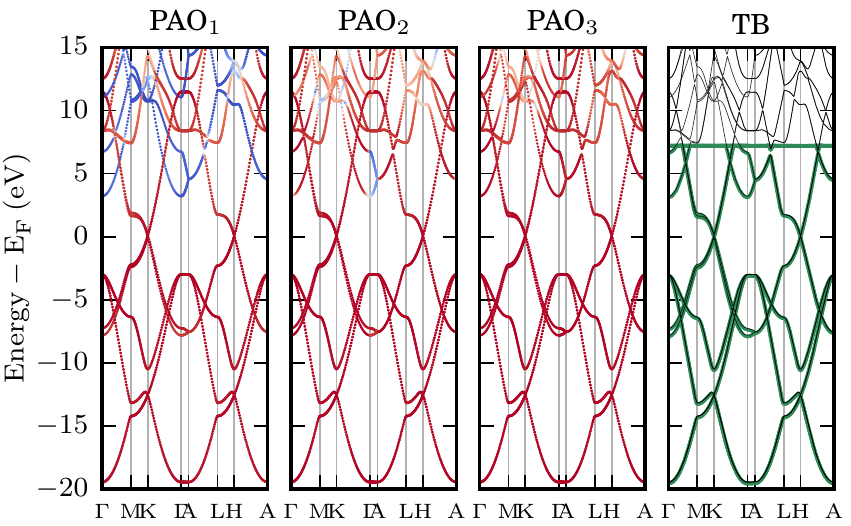}
\vspace{-5mm}
\caption{\small (Color online)  
Projectability $p_{n\bk}$, in color scale, of Bloch states $\psi_{n\bk}$ of graphite on the PAO$_1$, PAO$_2$, and PAO$_3$ sets. The fourth panel shows the interpolated band structure in green obtained from the TB Hamiltonians built on the PAO$_3$, superimposed to the plane-wave band structure. The color bar is shown in Fig.~\ref{fig:fig_graphene}. 
}
\label{fig:graphite} 
\end{center}
\end{figure}

The first step toward quantifying the quality of LCAO basis set involves a detailed analysis of the projectabilities. As mentioned before we construct PAO sets of increasing ``completeness'' and  derived an appropriate TB model intended to interpolate the fully converged PW band structure. In Fig.~\ref{fig:graphite} we discriminate between bands with moderate to high projectabilities (plotted in red) from bands with lower projectabilities (plotted in blue.) See color scale in Fig.~\ref{fig:fig_graphene}.  
The first panel shows the projectabilities on the PAO$_1$ set. There are two discernible groups of bands: the states with predominant  $\{s,p\}$ character (in red) and the parabolic bands near $\Gamma$ (in blue). Expectedly, the high projectabilitis bands qualitatively resemble the LCAO-DFT calculation with SZ basis in Fig.~\ref{fig:convergence}. Conversely, the low projectability bands (in blue), which can not be well represented on the minimal $\{s,p\}$ basis, are absent in Fig.~\ref{fig:convergence} (SZ.) Interestingly, those bands correspond to the so-called interlayer states of graphite\cite{Posternak1983PredictionInterlayerStates,Fischer1991Graphitic_interlayer_states,Csanyi2005GraphiteInterlayer}, which are strongly-dispersed unoccupied states located in-between the graphitic planes (see Fig.~3c in Ref.~\onlinecite{Csanyi2005GraphiteInterlayer}.)
 
Interlayer states are characteristic of materials with interstitial hollow regions,\cite{Matsushita2012FloatingStates,Matsushita2014InterstitialStates} layered structures,\cite{Blase1995Quasiparticle_BN} and/or reduced dimensionality such as carbon nanotubes,\cite{Okada2000CNT_interlayer,Chiou2002CNT_interlayer_states} C$_{60}$,\cite{Feng2008C60States} etc. Naturally, atom-centered basis functions are inappropriate to describe states that extend to the interstitial and/or vacuum regions, whereas PW basis are particularly well suited for this.\cite{Stewart2012Cautionary_tale} This is reflected in the pronounced discrepancy between unoccupied eigenenergies calculated using PW and AO basis sets.\cite{Matsushita2012FloatingStates,Matsushita2014InterstitialStates} The interlayer states of graphite are not captured with commonly used AO basis sets\cite{Samuelson1980Graphite,Trickey1992GrapheneDFT} such as DZP, as seen in Fig.~\ref{fig:convergence}. 

Our PAW-based procedure allows a systematic extension of the PAO sets (see Table~\ref{tab:table1}) to distill TB Hamiltonians that capture the details of the band structure including ``interlayer states''. The performance of each PAO set is assessed by determining its energy range of good projectability for a particular material, which in turn leads to a TB model that is highly accurate within the same energy range.

\begin{figure}[htb]
\begin{center}
\includegraphics[width=0.99\columnwidth]{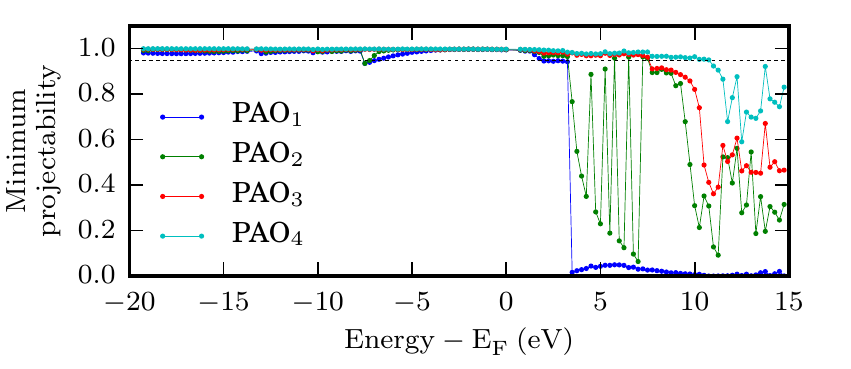}
\vspace{-5mm}
\caption{\small (Color online)
Projectability of the Bloch states $\psi_{n\bk}$ of graphite onto the four PAO sets defined in Table~\ref{tab:table1}. The plot shows the minimum value of projectability for each bin of a discretized energy grid. The dotted line indicates the projectability threshold of 0.95. All $\bk$ points in the reciprocal unit cell are included in the calculation.
}
\label{fig:projectability} 
\end{center}
\end{figure}

The performance of  the four PAO sets in Table~\ref{tab:table1} in terms of projectabilities are shown in Fig.~\ref{fig:projectability}. We choose the threshold of $p_{n\bk} \ge 0.95$ to determine the target energy range of accuracy.
The performance of the PAO$_1$ set (blue line)  sharply declines above 3.3 eV due to the presence of parabolic bands that do not project well on the small LCAO basis. The TB model using this set is accurate only up to 3.3 eV. The larger spatial range of the $\{s',p'\}$ functions added in PAO$_2$ facilitates the representation of the interlayer states closer to the graphitic planes, yielding a noticeable increase in the projectabilities with respect to PAO$_1$ over the entire energy range. This is also observed in the second panel in Fig.~\ref{fig:graphite} where most of the blue bands switched to red; the interlayer states with wave-vector component perpendicular to the graphitic planes, those in the $\Gamma$--A direction, are still of low projectability. The wavefunction of perpendicular Bloch states are primarily localized at the center of the interstitial space, the farthest from the planes, and the $\{s',p'\}$ functions are still insufficient to fully capture them; this is reflected by the oscillations of the green line in the 3.3--7 eV range in Fig.~\ref{fig:projectability}. 
When we augment the basis set with $d$ functions (PAO$_3$) the range of high projectability increases up to 7.2 eV above the Fermi level. PAO$_4$ (light blue line) further extends that range up to 10.8 eV. From the above results it is clear that the upper bound of the energy range of high-projectability, and consequently the range of accuracy of the TB matrices, can be systematically increased in this way.

The interpolated TB band structure constructed using the PAO$_3$ set is shown in green in Fig.~\ref{fig:graphite}. An excellent agreement with the PW-DFT bands up to 7.2 eV above the Fermi level is observed, as expected from the energy range of high projectabilities deduced from Fig.~\ref{fig:projectability}. 

\subsection{Graphene}

\begin{figure}[htb]
\begin{center}
\includegraphics[width=0.99\columnwidth]{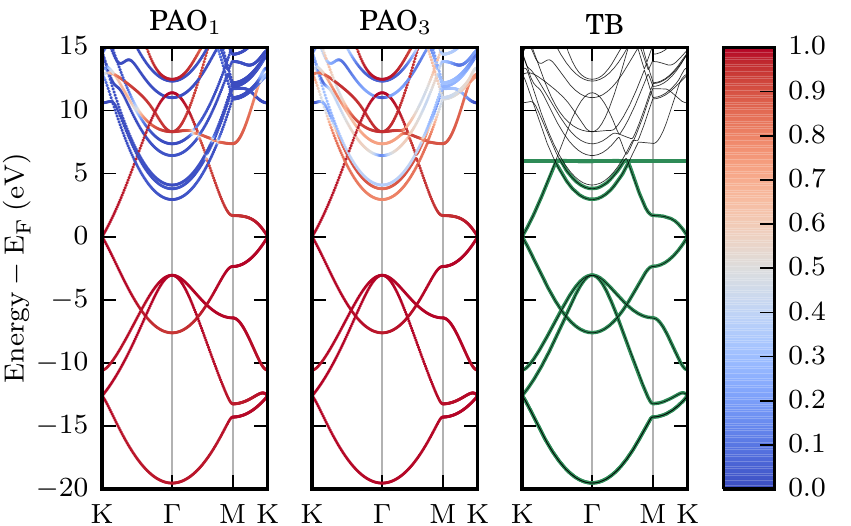}
\vspace{-5mm}
\caption{\small (Color online)
Projectabilities of the Bloch states of graphene on the PAO$_1$ and PAO$_3$ sets. The interpolated TB band structure, computed using the PAO$_3$ basis set, is shown in green on the third panel superimposed to the PW bands.}
\label{fig:fig_graphene} 
\end{center}
\end{figure}

The high energy electronic bands of graphene are characterized by the presence of image states.  
Image states give rise to superconductivity in metal-doped graphite.\cite{doped_graphite_CaYb,Csanyi2005GraphiteInterlayer} Similarly, they play a critical role in 
several phenomena such as the functionalization of graphene, the adsorption of oxygenated moieties and hydrogen,  the formation of defects, \cite{Coleman2008Graphene_defects,Ehlert2014Graphene_defects} and the `finger-print' peaks in the X-ray absorption spectra
found in the same energy region.\cite{Pacile2009Reply,Pacile2008Graphene_interlayer_xray,Papagno2009Graphene,Jeong2009Comment_interlayer_states,Schultz2011Imaging_graphene,Hua2010Theory_interlayer_states} Image states have also been shown to mediate electron tunneling in graphene.\cite{Zhang2008PhononGraphene} 
The importance of  TB models that reproduce well ``interlayer'' and ``image states'' cannot be underestimated especially when designing devices and interpreting experiments.

Image states in graphene follow\cite{Silkin2009GrapheneInterlayerStates} double Rydberg series $n^\pm$ and, expectedly, have low projectability on PAO$_1$ (blue lines in the  first panel of Fig.~\ref{fig:fig_graphene}). The two lowest unoccupied bands, parabolic at $\Gamma$, are the first states of the series, denoted as $1^+$ and $1^-$. The third lowest unoccupied band corresponds to the state $2^+$. 
Increasing the projectability of the image-state bands by augmenting the basis set is difficult for the particular case of graphene. The projectabilities improve with higher PAO sets, but fail to reach the threshold of 0.95. As seen in the second panel, bands $1^+$ and $1^-$ can reach moderately high projectability ($\sim 0.8$ at $\Gamma$) with PAO$_3$, in sharp contrast, however, band $2^+$ (light blue) still exhibits low projectability. This behavior is due to the spatial distribution of the image states. 
Indeed, the wavefunctions of $1^+$ and $1^-$ around $\Gamma$ have a component that is localized in the graphene plane forming $\sigma$ and $\pi$ hybridizations, respectively (see also Fig.~3b in Ref.~\onlinecite{Silkin2009GrapheneInterlayerStates}). 
This component can be partially accounted for with the inclusion of $\{s', p'\}$ (and to a lesser extent, $d$) functions in PAO$_3$, leading to the observed increase of projectabilities in the second panel with respect to the first in Fig.~\ref{fig:fig_graphene}. The other component of the wavefunctions $1^+$ and $1^-$ is more localized in the vacuum region\cite{Silkin2009GrapheneInterlayerStates} and, thus, is not captured by the $s'$ and $p'$ functions.
%
Band $1^-$ loses its dispersion and becomes flat around K, where a marked reduction in projectability in also seen (blue segment at $\sim 10.6$ eV in the second panel.) This happens because around K the wavefunction of $1^-$ loses its in-plane $\pi$-like component and consequently can no longer be expanded with the $\{s',p'\}$ functions (see also the charge density plot in Fig.~3a in Ref.~\onlinecite{Kogan2014GrapheneTB}). 
On the other hand, the wavefunction $2^+$ is fully localized in the vacuum region with the position of the maximum electron density away from the graphene plane\cite{Silkin2009GrapheneInterlayerStates}. Contrary to $1^\pm$, it has no in-plane component at $\Gamma$ and, thus, cannot be represented just by adding basis functions that are centered in the plane; therefore, $2^+$ exhibits low projectability on the PAO$_3$ set. 

This is corroborated by examining the LCAO-DFT bands, shown in the first three panels in Fig.~\ref{fig:fig_graphene1}. Even with the large TZDP basis set (second panel), the band structure completely misses the parabolic image state bands obtained when using the PW basis. The image states are reached only after extending the DZP with empty-atom (EA) basis functions centered at 2.8 \AA\ above and below the graphene sheet. Although the DZP-EA set is too expensive for practical calculations, it is observed that the bands (third panel) reach qualitative agreement to the PW solution (fourth panel). 

As seen in the third panel in Fig.~\ref{fig:fig_graphene}, the TB Hamiltonian generated using the PAO$_3$ set is able to correctly reproduce the band structure (in green) up to the two lowest parabolic bands $1^+$ and $1^-$. Our method supersedes parameterized tight-binding schemes that are only suitable in the vicinity of the Dirac point. The TB band structure shown here achieves higher accuracy than even the LCAO-DFT result, and with a less expensive basis set.
We expect that extending the PAO set with empty-atom functions located off plane will noticeably increase the projectabilities of the interlayer states $1^\pm$ and $2^+$ above the threshold of 0.95. We leave this for future investigation.

\section{Summary and Conclusions}
We presented a scheme to extend PAO basis sets to systematically increase the level of completeness of tight-binding representations obtained from plane waves {\it ab initio} calculations.
While minimal PAO sets (of SZ or SZP quality) can be sufficient for generating TB Hamiltonian matrices accurate up to $\sim$ 2 eV above the Fermi level for most materials, we have shown that enhanced PAO basis sets, containing both negative and positive energy atomic-orbital functions, can controllably increase the energy window in which the TB model is faithfully representing the details of the bands. 
Results for graphite and graphene, notably very difficult systems to represent in a TB scheme, demonstrate the accuracy and effectiveness of the method.

\ \\

\acknowledgments
We want to thank Dr. Derek Stewart for helpful discussions, the Texas Advanced Computing Center (TACC) at the University of Texas Austin for providing computing facilities, and the funding provided by the ONR-MURI under Contract No. N00014-13-1-0635. 
The authors acknowledge the Duke University Center for Materials Genomics and the CRAY Corporation for computational assistance.

\bibliographystyle{PhysRevwithTitles_noDOI_v1b}

\begin{thebibliography}{10}

\bibitem{Huckel1931Model}
E.~H{\"u}ckel, \emph{Quantentheoretische Beitr{\"a}ge zum Benzolproblem},
  Zeitschriftf{\"u}r Physik \textbf{70}, 628 (1931).

\bibitem{Jones1934Tight_binding}
H.~Jones, N.~F. Mott, and H.~W.~B. Skinner, \emph{A Theory of the Form of the
  X-Ray Emission Bands of Metals}, Phys.\ Rev. \textbf{45}, 379--384 (1934).

\bibitem{nmatHT}
S.~Curtarolo, G.~L.~W. Hart, M.~{Buongiorno~Nardelli}, N.~Mingo, S.~Sanvito,
  and O.~Levy, \emph{The high-throughput highway to computational materials
  design}, Nature\ Mater. \textbf{12}, 191--201 (2013).

\bibitem{Andersen_muffintin_PRB2000}
O.~K. Andersen and T.~{Saha-Dasgupta}, \emph{Muffin-tin orbitals of arbitrary
  order}, Phys.\ Rev.\ B \textbf{62}, R16219--R16222 (2000).

\bibitem{Marzari2012MLWF}
N.~Marzari, A.~A. Mostofi, J.~R. Yates, I.~Souza, and D.~Vanderbilt,
  \emph{Maximally localized Wannier functions: Theory and applications}, Rev.\
  Mod.\ Phys. \textbf{84}, 1419--1475 (2012).

\bibitem{Lu2004QUAMBO}
W.~C. Lu, C.~Z. Wang, T.~L. Chan, K.~Ruedenberg, and K.~M. Ho,
  \emph{Representation of electronic structures in crystals in terms of highly
  localized quasiatomic minimal basis orbitals}, Phys.\ Rev.\ B \textbf{70},
  041101 (2004).

\bibitem{Agapito_2013_projectionsPRB}
L.~A. Agapito, A.~Ferretti, A.~Calzolari, S.~Curtarolo, and
  M.~{Buongiorno~Nardelli}, \emph{Effective and accurate representation of
  extended Bloch states on finite Hilbert spaces}, Phys.\ Rev.\ B \textbf{88},
  165127 (2013).

\bibitem{Agapito2015TB_minimalPAO}
L.~A. Agapito, S.~{Ismail-Beigi}, S.~Curtarolo, M.~Fornari, and
  M.~{Buongiorno~Nardelli}, \emph{Accurate tight-binding Hamiltonian matrices
  from ab-initio calculations: Minimal basis sets}, Phys.\ Rev.\ B \textbf{93}, 035104 (2016).

\bibitem{curtarolo:art93}
L.~A. Agapito, S.~Curtarolo, and M.~{Buongiorno~Nardelli}, \emph{Reformulation
  of $\mathrm{DFT}+U$ as a Pseudohybrid Hubbard Density Functional for
  Accelerated Materials Discovery}, Phys.\ Rev.\ X \textbf{5}, 011006 (2015).

\bibitem{Pavan201259}
B.~Pavan, D.~Ceresoli, M.~M. Tecklenburg, and M.~Fornari, \emph{First
  principles NMR study of fluorapatite under pressure}, Solid State Nuclear
  Magnetic Resonance \textbf{45–46}, 59 -- 65 (2012).

\bibitem{PhysRevB.66.195107}
M.~Taillefumier, D.~Cabaret, A.-M. Flank, and F.~Mauri, \emph{X-ray absorption
  near-edge structure calculations with the pseudopotentials: Application to
  the \textit{K} edge in diamond and $\ensuremath{\alpha}$-quartz}, Phys. Rev.
  B \textbf{66}, 195107 (2002).

\bibitem{PhysRevB.53.10942}
A.~Pasquarello, M.~S. Hybertsen, and R.~Car, \emph{Theory of Si 2 \textit{p}
  core-level shifts at the Si(001)-${\mathrm{SiO}}_{2}$ interface}, Phys. Rev.
  B \textbf{53}, 10942--10950 (1996).

\bibitem{Bloechl1990GeneralizedPotentials}
P.~E. Bl\"ochl, \emph{Generalized separable potentials for electronic-structure
  calculations}, Phys.\ Rev.\ B \textbf{41}, 5414--5416 (1990).

\bibitem{DalCorso2014Pseudos}
A.~{Dal Corso}, \emph{Pseudopotentials periodic table: From H to Pu}, Comp.\
  Mat.\ Sci. \textbf{95}, 337 -- 350 (2014).

\bibitem{gbrv}
K.~F. Garrity, J.~W. Bennett, K.~M. Rabe, and D.~Vanderbilt,
  \emph{Pseudopotentials for high-throughput {DFT} calculations}, Comp.\ Mat.\
  Sci. \textbf{81}, 446--452 (2014).

\bibitem{Jollet2014PAWlib}
F.~Jollet, M.~Torrent, and N.~Holzwarth, \emph{Generation of Projector
  Augmented-Wave atomic data: A 71 element validated table in the XML format},
  Comput.\ Phys.\ Commun. \textbf{185}, 1246 -- 1254 (2014).

\bibitem{Note1}
\protect \texttt {https://wiki.fysik.dtu.dk/gpaw/setups/setups.html}.

\bibitem{kresse_vasp_paw}
G.~Kresse and D.~Joubert, \emph{From ultrasoft pseudopotentials to the
  projector augmented-wave method}, Phys.\ Rev.\ B \textbf{59}, 1758 (1999).

\bibitem{Note2}
\protect \texttt {http://users.wfu.edu/natalie/papers/pwpaw/newperiodictable/}.

\bibitem{Ceresoli2010GIPAW}
D.~Ceresoli, N.~Marzari, M.~G. Lopez, and T.~Thonhauser, \emph{\textit{Ab
  initio} converse NMR approach for pseudopotentials}, Phys.\ Rev.\ B
  \textbf{81}, 184424 (2010).

\bibitem{Kanwal2013linear}
R.~P. Kanwal, \emph{Linear integral equations} (Springer Science \& Business
  Media, 2013).

\bibitem{Hamann1989Generalized_NCPP}
D.~R. Hamann, \emph{Generalized norm-conserving pseudopotentials}, Phys.\ Rev.\
  B \textbf{40}, 2980--2987 (1989).

\bibitem{quantum_espresso_2009}
P.~Giannozzi, S.~Baroni, N.~Bonini, M.~Calandra, R.~Car, C.~Cavazzoni,
  D.~Ceresoli, G.~L. Chiarotti, M.~Cococcioni, I.~Dabo, A.~{Dal Corso}, S.~{de
  Gironcoli}, S.~Fabris, G.~Fratesi, R.~Gebauer, U.~Gerstmann, C.~Gougoussis,
  A.~Kokalj, M.~Lazzeri, L.~Martin-Samos, N.~Marzari, F.~Mauri, R.~Mazzarello,
  S.~Paolini, A.~Pasquarello, L.~Paulatto, C.~Sbraccia, S.~Scandolo,
  G.~Sclauzero, A.~P. Seitsonen, A.~Smogunov, P.~Umari, and R.~M. Wentzcovitch,
  \emph{QUANTUM ESPRESSO: a modular and open-source software project for
  quantum simulations of materials}, J.\ Phys.:\ Conden.\ Matt. \textbf{21},
  395502 (2009).

\bibitem{Ozaki2003BasisFunctions}
T.~Ozaki, \emph{Variationally optimized atomic orbitals for large-scale
  electronic structures}, Phys.\ Rev.\ B \textbf{67}, 155108 (2003).

\bibitem{PBE}
J.~P. Perdew, K.~Burke, and M.~Ernzerhof, \emph{Generalized gradient
  approximation made simple}, Phys.\ Rev.\ Lett. \textbf{77}, 3865--3868
  (1996).

\bibitem{Ozaki2004Adpack}
T.~Ozaki and H.~Kino, \emph{Numerical atomic basis orbitals from H to Kr},
  Phys.\ Rev.\ B \textbf{69}, 195113 (2004).

\bibitem{Gusso2008LCAOvsPW}
M.~Gusso, \emph{Study on the maximum accuracy of the pseudopotential density
  functional method with localized atomic orbitals versus plane-wave basis
  sets}, J.\ Chem.\ Phys. \textbf{128}, 044102 (2008).

\bibitem{Gross1996TDDFT}
M.~Petersilka, U.~J. Gossmann, and E.~K.~U. Gross, \emph{Excitation Energies
  from Time-Dependent Density-Functional Theory}, Phys.\ Rev.\ Lett.
  \textbf{76}, 1212--1215 (1996).

\bibitem{Gorling1995DFPT}
A.~G{\"o}rling and M.~Levy, \emph{DFT ionization formulas and a DFT
  perturbation theory for exchange and correlation, through adiabatic
  connection}, International Journal of Quantum Chemistry \textbf{56}, 93--108
  (1995).

\bibitem{Hybertsen1986GW}
M.~S. Hybertsen and S.~G. Louie, \emph{Electron correlation in semiconductors
  and insulators: Band gaps and quasiparticle energies}, Phys.\ Rev.\ B
  \textbf{34}, 5390--5413 (1986).

\bibitem{Taube2008CCbasis}
A.~G. Taube and R.~J. Bartlett, \emph{Frozen natural orbital coupled-cluster
  theory: Forces and application to decomposition of nitroethane}, J.\ Chem.\
  Phys. \textbf{128}, 164101 (2008).

\bibitem{Posternak1983PredictionInterlayerStates}
M.~Posternak, A.~Baldereschi, A.~J. Freeman, E.~Wimmer, and M.~Weinert,
  \emph{Prediction of Electronic Interlayer States in Graphite and
  Reinterpretation of Alkali Bands in Graphite Intercalation Compounds}, Phys.\
  Rev.\ Lett. \textbf{50}, 761--764 (1983).

\bibitem{Fischer1991Graphitic_interlayer_states}
D.~A. Fischer, R.~M. Wentzcovitch, R.~G. Carr, A.~Continenza, and A.~J.
  Freeman, \emph{Graphitic interlayer states: A carbon \textit{K} near-edge
  x-ray-absorption fine-structure study}, Phys.\ Rev.\ B \textbf{44},
  1427--1429 (1991).

\bibitem{Csanyi2005GraphiteInterlayer}
G.~Cs{\'a}nyi, P.~Littlewood, A.~H. Nevidomskyy, C.~J. Pickard, and B.~Simons,
  \emph{The role of the interlayer state in the electronic structure of
  superconducting graphite intercalated compounds}, Nature\ Phys. \textbf{1},
  42--45 (2005).

\bibitem{Matsushita2012FloatingStates}
Y.-i. Matsushita, S.~Furuya, and A.~Oshiyama, \emph{Floating Electron States in
  Covalent Semiconductors}, Phys.\ Rev.\ Lett. \textbf{108}, 246404 (2012).

\bibitem{Matsushita2014InterstitialStates}
Y.-i. Matsushita and A.~Oshiyama, \emph{Interstitial Channels that Control Band
  Gaps and Effective Masses in Tetrahedrally Bonded Semiconductors}, Phys.\
  Rev.\ Lett. \textbf{112}, 136403 (2014).

\bibitem{Blase1995Quasiparticle_BN}
X.~Blase, A.~Rubio, S.~G. Louie, and M.~L. Cohen, \emph{Quasiparticle band
  structure of bulk hexagonal boron nitride and related systems}, Phys.\ Rev.\
  B \textbf{51}, 6868--6875 (1995).

\bibitem{Okada2000CNT_interlayer}
S.~Okada, A.~Oshiyama, and S.~Saito, \emph{Nearly free electron states in
  carbon nanotube bundles}, Phys.\ Rev.\ B \textbf{62}, 7634--7638 (2000).

\bibitem{Chiou2002CNT_interlayer_states}
J.~W. Chiou, C.~L. Yueh, J.~C. Jan, H.~M. Tsai, W.~F. Pong, I.-H. Hong,
  R.~Klauser, M.-H. Tsai, Y.~K. Chang, Y.~Y. Chen, C.~T. Wu, K.~H. Chen, S.~L.
  Wei, C.~Y. Wen, L.~C. Chen, and T.~J. Chuang, \emph{Electronic structure of
  the carbon nanotube tips studied by x-ray-absorption spectroscopy and
  scanning photoelectron microscopy}, Appl.\ Phys.\ Lett. \textbf{81},
  4189--4191 (2002).

\bibitem{Feng2008C60States}
M.~Feng, J.~Zhao, and H.~Petek, \emph{Atomlike, Hollow-Core--Bound Molecular
  Orbitals of C60}, Science \textbf{320}, 359--362 (2008).

\bibitem{Stewart2012Cautionary_tale}
D.~Stewart, \emph{A Cautionary Tale of Two Basis Sets and Graphene}, Computing
  in Science and Engineering \textbf{14}, 55--59 (2012).

\bibitem{Samuelson1980Graphite}
L.~Samuelson and I.~P. Batra, \emph{Electronic properties of various stages of
  lithium intercalated graphite}, J.\ Phys.\ C:\ Solid\ State\ Phys.
  \textbf{13}, 5105 (1980).

\bibitem{Trickey1992GrapheneDFT}
S.~B. Trickey, F.~M\"uller-Plathe, G.~H.~F. Diercksen, and J.~C. Boettger,
  \emph{Interplanar binding and lattice relaxation in a graphite dilayer},
  Phys.\ Rev.\ B \textbf{45}, 4460--4468 (1992).

\bibitem{doped_graphite_CaYb}
T.~E. Weller, M.~Ellerby, S.~S. Saxena, R.~P. Smith, and N.~T. Skipper,
  \emph{Superconductivity in the intercalated graphite compounds C$_6$Yb and
  C$_6$Ca}, Naturere Physics \textbf{1}, 39--41 (2005).

\bibitem{Coleman2008Graphene_defects}
V.~A. Coleman, R.~Knut, O.~Karis, H.~Grennberg, U.~Jansson, R.~Quinlan, B.~C.
  Holloway, B.~Sanyal, and O.~Eriksson, \emph{Defect formation in graphene
  nanosheets by acid treatment: an x-ray absorption spectroscopy and density
  functional theory study}, J.\ Phys.\ Chem. \textbf{41}, 062001 (2008).

\bibitem{Ehlert2014Graphene_defects}
C.~Ehlert, W.~E.~S. Unger, and P.~Saalfrank, \emph{C K-edge NEXAFS spectra of
  graphene with physical and chemical defects: a study based on density
  functional theory}, Phys.\ Chem.\ Chem.\ Phys. \textbf{16}, 14083--14095
  (2014).

\bibitem{Pacile2009Reply}
D.~Pacil\'e, M.~Papagno, A.~F. Rodr\'{i}guez, M.~Grioni, L.~Papagno, {\c
  C}.~Girit, J.~C. Meyer, G.~E. Begtrup, and A.~Zettl, \emph{Pacil\'e
  \textit{et al.} Reply:}, Phys.\ Rev.\ Lett. \textbf{102}, 099702 (2009).

\bibitem{Pacile2008Graphene_interlayer_xray}
D.~Pacil\'e, M.~Papagno, A.~F. Rodr\'{i}guez, M.~Grioni, L.~Papagno, {\c
  C}.~Girit, J.~C. Meyer, G.~E. Begtrup, and A.~Zettl, \emph{Near-Edge X-Ray
  Absorption Fine-Structure Investigation of Graphene}, Phys.\ Rev.\ Lett.
  \textbf{101}, 066806 (2008).

\bibitem{Papagno2009Graphene}
M.~Papagno, A.~F. Rodr{\'\i}guez, {\c C}.~Girit, J.~Meyer, A.~Zettl, and
  D.~Pacil{\'e}, \emph{Polarization-dependent C K near-edge X-ray absorption
  fine-structure of graphene}, Chem.\ Phys.\ Lett. \textbf{475}, 269 -- 271
  (2009).

\bibitem{Jeong2009Comment_interlayer_states}
H.-K. Jeong, H.-J. Noh, J.-Y. Kim, L.~Colakerol, P.-A. Glans, M.~H. Jin, K.~E.
  Smith, and Y.~H. Lee, \emph{Comment on ``Near-Edge X-Ray Absorption
  Fine-Structure Investigation of Graphene''}, Phys.\ Rev.\ Lett. \textbf{102},
  099701 (2009).

\bibitem{Schultz2011Imaging_graphene}
B.~J. Schultz, C.~J. Patridge, V.~Lee, C.~Jaye, P.~S. Lysaght, C.~Smith,
  J.~Barnett, D.~A. Fischer, D.~Prendergast, and S.~Banerjee, \emph{Imaging
  local electronic corrugations and doped regions in graphene}, Nature\ Commun.
  \textbf{2}, 372 (2011).

\bibitem{Hua2010Theory_interlayer_states}
W.~Hua, B.~Gao, S.~Li, H.~\AA{}gren, and Y.~Luo, \emph{X-ray absorption spectra
  of graphene from first-principles simulations}, Phys.\ Rev.\ B \textbf{82},
  155433 (2010).

\bibitem{Zhang2008PhononGraphene}
Y.~Zhang, V.~W. Brar, F.~Wang, C.~Girit, Y.~Yayon, M.~Panlasigui, A.~Zettl, and
  M.~F. Crommie, \emph{Giant phonon-induced conductance in scanning tunnelling
  spectroscopy of gate-tunable graphene}, Nature\ Phys. \textbf{4}, 627--630
  (2008).

\bibitem{Silkin2009GrapheneInterlayerStates}
V.~M. Silkin, J.~Zhao, F.~Guinea, E.~V. Chulkov, P.~M. Echenique, and H.~Petek,
  \emph{Image potential states in graphene}, Phys.\ Rev.\ B \textbf{80}, 121408
  (2009).

\bibitem{Kogan2014GrapheneTB}
E.~Kogan, V.~U. Nazarov, V.~M. Silkin, and M.~Kaveh, \emph{Energy bands in
  graphene: Comparison between the tight-binding model and \textit{ab initio}
  calculations}, Phys.\ Rev.\ B \textbf{89}, 165430 (2014).

\end{thebibliography}

\end{document}